\documentclass[aps,prx,twocolumn,superscriptaddress]{revtex4-1}
\UseRawInputEncoding
\usepackage{multirow}
\usepackage{latexsym}
\usepackage{amssymb}
\usepackage{upgreek}
\usepackage{graphicx,epstopdf}
\usepackage{mathrsfs}
\usepackage{dcolumn}
\usepackage{bm,dsfont}
\usepackage{amsmath}
\usepackage{amsfonts}
\usepackage{color}
\usepackage{hyperref}
\hypersetup{
colorlinks = true,
linkcolor = [rgb]{0,0,1},
citecolor = [rgb]{0.75,0,0},
urlcolor = [rgb]{0.25, 0.21, 1}}
\DeclareMathOperator{\sgn}{sgn}
\newcommand{\vect}[1]{{\bm{#1}}}
\newcommand{\bra}[1]{\langle #1|}
\newcommand{\ket}[1]{|#1 \rangle}
\newcommand{\kp}{\vect{k}^\prime}
\newcommand{\eq}[1]{\begin{equation}#1\end{equation}}
\newcommand{\eqnref}[1]{Eq.\,\eqref{#1}}
\newcommand{\figref}[1]{Fig.\,\ref{#1}}
\usepackage{xcolor}
\usepackage{mathrsfs}

\begin{document}
\title{Topology in non-Hermitian Chern insulators with skin effect}
\author{Yi-Xin Xiao}
\affiliation{Department of Physics, Hong Kong University of Science and Technology, Clear Water Bay, Hong Kong, China}
\author{C. T. Chan}
\email[Electronic address: ]{phchan@ust.hk}
\affiliation{Department of Physics, Hong Kong University of Science and Technology, Clear Water Bay, Hong Kong, China}

\begin{abstract}
The non-Hermitian skin effect can arise in materials that have asymmetric hoppings between atoms or resonating units, which makes the bulk eigen-spectrum sensitive to boundary conditions. When skin effect emerges, eigenstates in the bulk continuum can become localized on the edges, making the distinction between edge and bulk states challenging. We establish the bulk-boundary correspondence for a Chern insulator model with non-Hermitian skin effect by combining two approaches (``non-Bloch'' approach and ``biorthogonal'' approach). Both approaches can suppress the skin effect but they are based on different mathematical tools. A biorthogonal inverse participation ratio is used as a measure to distinguish between bulk states and edge states, and a non-Bloch Chern number is used to characterize the topology and predict the number of topological edge bands. In addition to tangential degeneracies, crossing degeneracies are found to occur between the bulk and edge bands. Their presence enriches the (de)localization behavior of the edge states but does not affect the Chern number. The phase diagram of the system has interesting features that are not found in Hermitian systems. For example, one topological transition and two non-Hermitian phase transitions can be induced by tuning a single parameter. The gapless phase is topologically protected due to the stable existence of the non-Hermitian band degeneracies guaranteed by nonzero discriminant numbers. 
\end{abstract}
\maketitle
\section*{I. Introduction}
Topological band theory has inspired extensive research in various domains of physics and material science \cite{hasan2010,qi2011RMP,bansil2016}. Among various interesting implications of topological theory, the bulk-boundary correspondence (BBC) is regarded as a cornerstone as it relates the bulk topological invariant \cite{Bernevig2013,chiu2016} to topological edge modes, which have many potential applications due to their robustness. The BBC is well established for Hermitian Hamiltonians. 

Most real-world systems are coupled to their environment and thus cannot be completely described by Hermitian Hamiltonians. Non-Hermitian Hamiltonians are needed for describing many physical systems such as wave systems with loss and gain \cite{el-ganainyReview2018,ozdemir2019PT,fengPhotonics2017,tang2020Nexus}, open systems \cite{rotter2009open,lee2014,zhen2015Spawning,cao2015} and solids where non-Hermitian self-energy emerges due to electron interactions \cite{Yoshida2018heavy,Yoshida2019correlated} or disorders \cite{shen2018Oscillation,papaj2019Nodal}. Diverse phenomena in non-Hermitian systems and especially those involving exceptional points (EPs) have attracted booming interest \cite{el-ganainyReview2018,ozdemir2019PT,fengPhotonics2017,xiao2019AnisotropicEP}. 

In particular, the non-Hermitian skin effect (NHSE) is a remarkable phenomenon that can emerge when the coupling parameters in a Hamiltonian become asymmetric, leading to the anomalous localization of bulk eigenstates at the boundaries \cite{lee2016,xiong2018Why,martinez2018,Frank2021dislocation}. The NHSE manifests itself as extreme sensitivity to boundary conditions, namely that the eigenspectrum of the open boundary conditions (OBC) dramatically differs from that calculated using periodic boundary conditions (PBC) \cite{lee2016,xiong2018Why,martinez2018}. Consequently the topological transitions in such OBC systems with NHSE cannot be accounted for properly by the conventional BBC that relies on the topological invariant of the periodic bulk \cite{yao2018Edge,yao2018Chern,kunstBio2018}. 

Topological properties of non-Hermitian systems have been studied \cite{imura2019,leeAnatomy2019,bergholtzRMP2021,yang2020AGBZ,yokomizo2019NonBloch,tang2020Nexus,
song2019Real,borgnia2020,shen2018Topological,gong2018PRX,kawabata2019PRX,kunst2019Transfer,
xiao2020BBC,zhou2019PeriodicTable,wang2021Science,kawabata2019classification,kawabata2021Field,
zhang2020Correspondence,okuma2020Origin,edvardsson2019,xiao2020Lieb,yang2021Doubling,
Xu2017WeylEP,Delplace2021symmetry,Mandal2021symmetry,Vecsei2021indicators} and it was uncovered that the NHSE originates from the intrinsic non-Hermitian topology \cite{kawabata2021Field,zhang2020Correspondence,okuma2020Origin}. Several works reestablished the BBC in the presence of NHSE \cite{yao2018Edge,yao2018Chern,kunstBio2018,imura2019, leeAnatomy2019, bergholtzRMP2021, yang2020AGBZ,yokomizo2019NonBloch,song2019Real,borgnia2020}. One approach is the non-Bloch approach that defines a topological invariant for a non-Bloch Hamiltonian constructed via the generalized Brillouin zone (GBZ), which embodies information of the OBC system \cite{yao2018Edge,yao2018Chern,yang2020AGBZ,yokomizo2019NonBloch}. Another approach is the biorthogonal approach, which treats a two-dimensional (2D) Chern insulator ribbon as a family of one-dimensional (1D) systems parameterized by $k_{\parallel}$ and directly relates topological phase transitions to jumps of biorthogonal polarization of boundary states, without involving bulk eigenvectors and Chern numbers \cite{kunstBio2018}. 

In this work, we consider the BBC in a certain class of boundary condition sensitive systems which has the skin effect due to asymmetric hoppings between resonating units or atoms. We provide a unified understanding of the BBC in such non-Hermitian systems, focusing on a prototypical system. The key lies in suppressing the NHSE so that the BBC could be treated in a similar way as Hermitian cases. We show that the bulk and edge states can be identified without ambiguity by using biorthogonal inverse participation ratio (bi-IPR) defined via the biorthogonal density \cite{gong2018PRX, tang2020Anderson,wang2019Transition} instead of wavefunctions. The edge states become delocalized when they touch the bulk bands. There are generally two types of such bulk-edge degeneracies: the tangential and the anomalous crossing degeneracies that have a non-Hermitian origin. The number of topological edge bands is dictated by the non-Bloch Chern numbers computed from the non-Bloch Hamiltonian. 

The phase diagram exhibits novel features, including the existence of both topological and non-Hermitian phase transitions \cite{wang2019Transition}, which are associated with Dirac points and EPs, respectively, as band degeneracies. The gapless phase is found to be topologically protected due to the topological protection of the non-Hermitian degeneracies (i.e., EPs) by nonzero discriminant numbers \cite{yang2021Doubling,xiao2020Lieb,shen2018Topological}.
\begin{figure}[htbp]
\includegraphics[clip, width=0.85\columnwidth, angle=0]{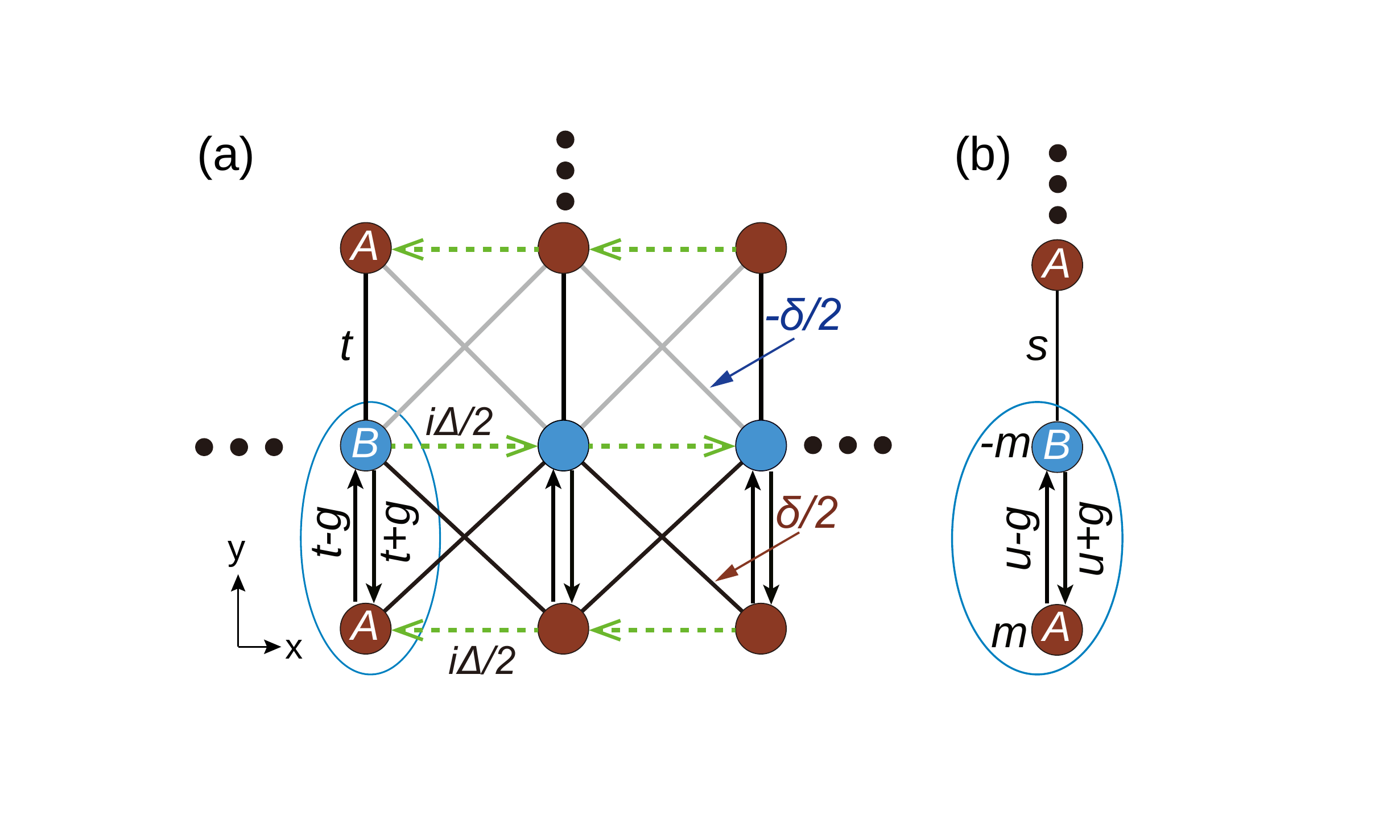}
\caption{(a) The non-Hermitian Chern insulator with two sites per unit cell. The dashed arrows denote imaginary hoppings. (b) The effective 1D chain model of (a). 
\label{Fig1}}
\end{figure}
\begin{figure}[htbp]
\includegraphics[clip, width=\columnwidth, angle=0]{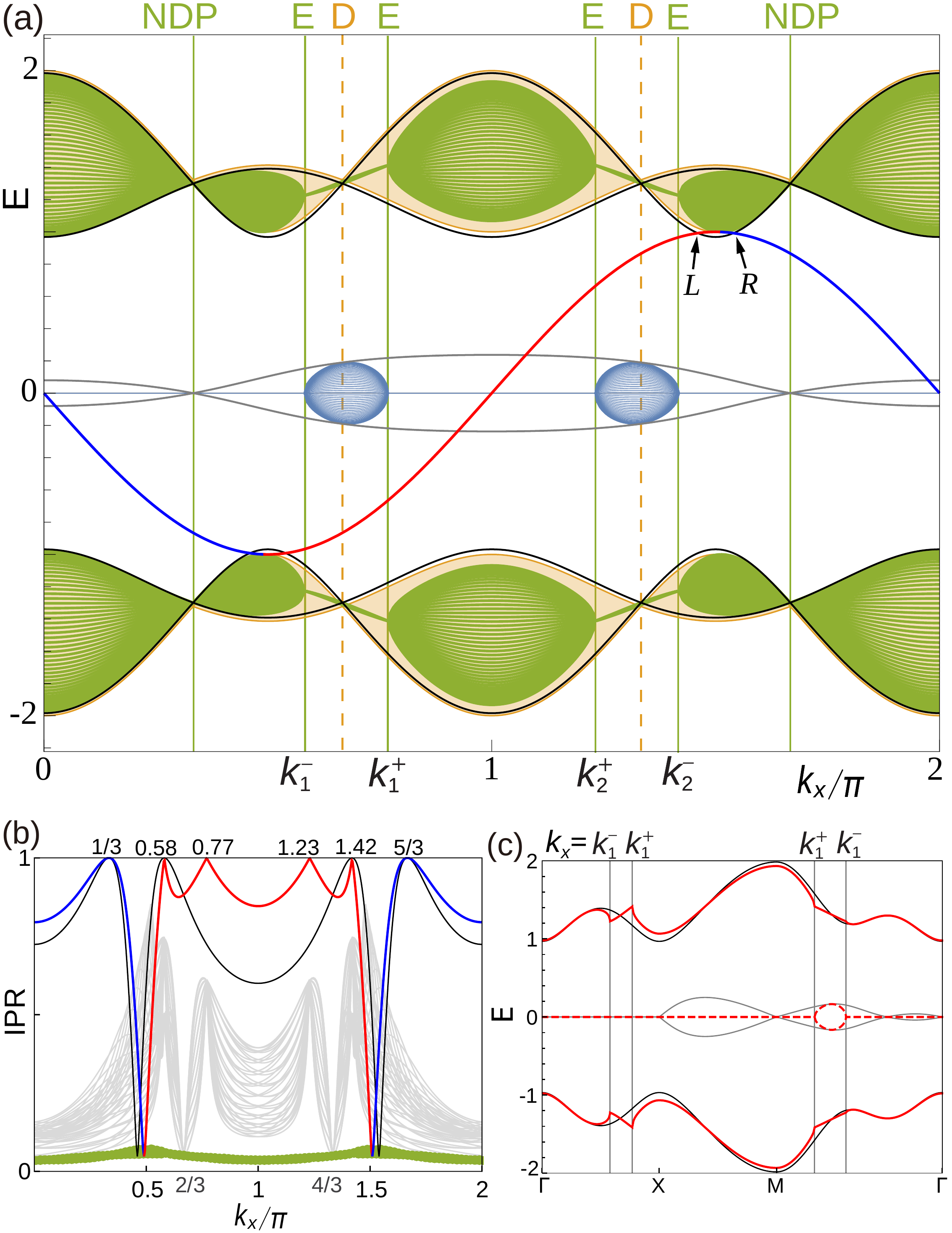}
\caption{(a) The band structure of an OBC ribbon with $N_y=101$ consists of green (navy) curves for the real (imaginary) part of bulk bands, and a blue/red curve for the edge band. The parameters are $\delta=\Delta=1$,$t=0.5$ and $g=0.25$. The orange bands are for $g=0$. The black (gray) curves denote the real (imaginary) part of the band edges of the PBC system. (b) The bi-IPR of the bulk (green) and edge (blue/red) states. The conventional IPRs of the bulk (gray) and edge (black) states are also shown. (c) The non-Bloch bands are denoted by red curves, with solid (dashed) curves for the real (imaginary) parts. The black (gray) curves denote the real (imaginary) parts of Bloch bands.
\label{Fig2}}
\end{figure}
\section*{II. Topology of Non-Hermitian Chern insulators with skin effect} 
\subsection*{A. Model system and Non-Hermitian skin effect} 
As a prototypical system exhibiting NHSE, we consider a non-Hermitian Chern insulator \cite{kunstBio2018} as shown pictorially in \figref{Fig1}(a). There are two sites ({\it A}, {\it B}) per unit cell and the hopping parameters $t, \Delta, \delta, g$ are all real numbers. The system contains imaginary hoppings $\pm i\Delta/2$ that breaks time-reversal symmetry, and asymmetric hoppings $t_{\downarrow / \uparrow}=t {\pm} g$ that cause NHSE in the y direction. The Bloch Hamiltonian takes the form, 
\eq{\label{eq:Hk}
H(\vect{k}) =
\begin{pmatrix}
m & u_+ + s e^{-i k_y} \\
u_- + s e^{i k_y} & -m
\end{pmatrix}
,} 
where $m=-\Delta \sin{k_x}$, $s=t-\delta \cos{k_x}$ and $u_{\pm} = u \pm g$ with $u=t+\delta \cos{k_x}$. For a given $k_x$, \eqnref{eq:Hk} describes the 1D Rice-Mele model \cite{rice1982} with asymmetric intracell hoppings, i.e., $u_- \ne u_+$ when $g\ne 0$, as shown in \figref{Fig1}(b). 

We consider a ribbon with $N_y=2n_y+1$ sites in the $y$ direction and PBC in the $x$ direction. Without loss of generality, we fix $\Delta=1$ and $t=0.5$, and allow $g$ and $\delta$ to vary. 

When $g=0$, the system is Hermitian and has a topologically nontrivial gap with gap size $E_g=4\delta$ for $\delta<t$ and $E_g=4t$ for $\delta>t$. In \figref{Fig2}(a), we show the ribbon's band structure, which comprises bulk bands (orange) and a topological edge band $E=m=-\Delta \sin{k_x}$ (red/blue), assuming $n_y=50$ and $N_y=101$. The edge band touches the orange bulk bands tangentially at $k_x=\pi/2$ and $3\pi/2$. For an odd $N_y$, the edge band $E=-\Delta \sin{k_x}$ always exists and the associated eigenvectors have analytic forms that vanish at all the {\it B} sites, reminiscent of the zero energy states in systems with sublattice (chiral) symmetry. There are four $n_y$-fold degeneracies at $k_x=2\pi/3$ and $4\pi/3$ where the intracell hoppings $u$ vanish, namely $t+\delta\cos{k_x}=0$, as marked by the orange gridlines labeled ``D''. 

When $g=0.25$, some eigenvalues become complex and the bulk bands are denoted by green and navy curves for $Re(E)$ and $Im(E)$, respectively. The edge band remains the same red/blue curve $E=-\Delta \sin{k_x}$ as the $g=0$ case. However, the bulk-edge degeneracies are shifted to $k_x \approx 0.49 \pi$ and $1.51\pi$, which divide the red and blue segments. The non-Hermitian system inherited the topological edge band from its Hermitian counterpart and is thus supposed to bear a BBC description. 

The non-Hermiticity ($g\ne 0$) splits the four $n_y$-fold degeneracies (for $g=0$) into four pairs of $n_y$-th-order EPs, as marked by green gridlines labeled ``E''. The EPs are located at $k_1^+ \approx 0.77\pi$, $k_2^+ \approx 1.23\pi$ and $k_{-1}^- \approx 0.58\pi$, $k_2^-\approx 1.42\pi$ for $g=0.25$, which are given by $u_+=u+g=0$ and $u_-=u-g=0$, respectively, that signify the vanishing of effective intracell hoppings. The spacing between each EP pair increases with $g$. In addition to the $n_y$-th-order EPs, there also exists another type of $n_y$-fold degeneracies at $k_x=\pi/3$ and $5\pi/3$ (labeled ``NDP'') where the intercell hopping vanish, i.e., $s=t-\delta \cos{k_x}=0$. In this case, the 1D chain is decoupled into $n_y $ identical dimers giving rise to $n_y$-fold non-defective degeneracy points (NDPs) at $E=\pm 3\sqrt{3}/4$. 

The PBC spectrum $\sigma(PBC)$ for $g=0.25$ is also shown in \figref{Fig2}(a), where its real and imaginary parts occupy the regions bounded by black curves and gray curves, respectively. The Bloch bands of \eqnref{eq:Hk} are given by $E_\pm(\vect{k})=\pm\sqrt{A+G}$ where $A=m^2+u^2+s^2+2us \cos{k_y}$ and $G=-g^2+2igs\sin{k_y}$ results from the non-Hermiticity, which accounts for the small perturbation of $O(g^2)$ in $Re[\sigma(PBC)]$ relative to the orange bands (Hermitian spectra), as shown in \figref{Fig2}(a). 

For an edge state with $E=m=-\Delta\sin{k_x}$, the right and left eigenvectors are 
\eq{\label{edge state}
\begin{split}
\ket{\psi}&=\left[1,\, 0,\, r,\, 0, \, r^2,  \cdots, 0,\, r^{n_y}\right]^T, \\
\bra{\varphi}&=\left[1,\, 0,\, l,\, 0,\, l^2,  \cdots, 0,\, l^{n_y}\right],
\end{split}
}
where $r=-u_-/s$ and $l= -u_+/s$. 

The right eigenvector $\ket{\psi}$ is delocalized when $u_-=s$, giving $k_x\approx 0.46\pi$ and $k_x\approx 1.54\pi$ (labeled {\it{``R''}} in \figref{Fig2}(a)), due to the balanced competition between the two localization mechanisms: the bandgap and the NHSE \cite{Zhu2021Delocalization}. For example near $k_x\approx 1.54\pi$, $\ket{\psi}$ is localized at the upper boundary when $g=0$, while the presence of NHSE ($g\ne 0$) tend to localize the states to the lower boundary due to $|u_+|>|u_-|$. It turns out that the delocalization points of $\ket{\psi}$ coincide with the crossing between the edge band and the PBC spectrum $\sigma(PBC)$, as is shown in \figref{Fig2}(a) where {\it{``R''}} lies at the intersection of the blue curve and the black curve that denotes the PBC band edge \cite{Zhu2021Delocalization}.

Furthermore, we found that the delocalization points of $\bra{\varphi}$, $k_x\approx 0.54\pi$ and $k_x\approx 1.46\pi$  (labeled {\it{``L''}}) given by $u_+=s$, also signify the degeneracies between the edge band and the PBC spectrum, as is shown in \figref{Fig2}(a). Such a result can be understood via the delocalization of the right eigenvector $\ket{\varphi}^\ast$ of $H^T$ noting that $H^T\ket{\varphi}^\ast=E\ket{\varphi}^\ast$ and $H^T$ has the same spectrum as $H$.

We note that $\ket{\psi}$ in \eqnref{edge state} is also delocalized when $u_-=- s$, leading to the condition $g=2t=1$. The $k_x$ independence implies that $\ket{\psi}$ is delocalized for all $k_x$. This corresponds to the special case when the PBC spectrum is gapless \cite{SeeSupp} and contains an eigenvalue $E(\vect{k})=-\Delta\sin k_x$ (when $k_y=0$) that is degenerate with the edge band of the OBC system which is gapped. Similarly $\bra{\varphi}$ is delocalized for all $k_x$ when $u_+=- s$.

The OBC and PBC spectra have striking differences for $g=0.25$, manifesting NHSE. Figure \ref{Fig2}(a) shows that the PBC bulk spectrum encloses the OBC bulk spectrum, like the case of one-band models studied in Refs. \cite{okuma2020Origin,wang2019Transition}. Due to NHSE, the wavefunctions of the OBC bulk states are localized, making it difficult to distinguish between bulk and edge states. We show that the bulk and edge states can be distinguished unambiguously using the biorthogonal IPR and the topological edge bands can be related to Chern numbers defined over the GBZ. 
\subsection*{B. Bulk-boundary correspondence} 
For a given state, its biorthogonal inverse participation ratio (bi-IPR) is defined as 
\eq{
I_{bi}=\sum\nolimits_{j=1}^{N_y} \rho_j^2 / \left(\sum\nolimits_{j=1}^{N_y} \rho_j\right)^2,
\label{bi-IPR}
}
where $\rho_j=|\varphi_j \psi_j |$ denotes the biorthogonal density at the $j$th site and involves both the left eigenvector $\bra{\varphi}$ and right eigenvector $\ket{\psi}$. The biorthogonal density has been demonstrated to be a good measure of localization in the study of Hatano-Nelson model \cite{gong2018PRX, hatano1998delocation}. We found that the topological edge states can be well distinguished from the skin-localized bulk states by the bi-IPR, since $|\varphi_j \psi_j|$ captures the delocalization property of bulk states. 

Specifically, the bi-IPRs tend to be small for all bulk states of the OBC system as denoted by the green curves in \figref{Fig2}(b). However, the bi-IPRs of the edge states are not small as denoted by the red and blue curves, which represent localization of biorthogonal density at the lower and upper boundaries, respectively. They are of order one except around the two delocalization points where the edge modes are degenerate with the bulk continuum.

For an edge state with eigenvectors given in \eqnref{edge state}, its biorthogonal density $\rho$ is 
\eq{\rho=|\varphi \circ \psi|=\mathcal{N}\left[1, 0,\mu, 0, \mu^2, \cdots,  0, \mu^{n_y} \right],
\label{rho}
}
where $\varphi \circ \psi$ means the component-wise product, $\mu=\left|u_- u_+/s^2\right|$, and
$\mathcal{N}$ is the biorthogonal normalization factor.
Requiring $\rho$ in \eqnref{rho} to be delocalized, the condition for the transition points is obtained to be \cite{SeeSupp}
\eq{u_+ u_-=\pm s^2.
\label{delocalize}
}

For the case of $\delta=1$ and $g=0.25$, the two delocalization points in \figref{Fig2}(b) are found from $u_+ u_-=s^2$ (i.e., $\cos{k_x}=g^2/4t\delta$) to be $k_x \approx 0.49\pi$ and $1.51\pi$, which coincide with the two tangential bulk-edge degeneracies (between the edge band and green OBC bulk bands) shown in \figref{Fig2}(a). The bi-IPRs are thus justified. 
As expected, the delocalization point $k_x\approx 1.51\pi$ of $\rho_e$ lies between {\it{``L''}} and {\it{``R''}}, the delocalization points of $\bra{\varphi}$ and $\ket{\psi}$, shown in \figref{Fig2}(a).

We find that $u_+ u_-=-s^2$ also predicts delocalization points of edge states, but they correspond to an unusual type of degeneracies that involve the crossing between the edge and bulk bands, which would be expounded later. Equation \eqref{delocalize} resembles the condition given by the non-Bloch approach for the topological transition in the non-reciprocal SSH model \cite{yao2018Edge}, indicating the equivalence of the two approaches in suppressing the NHSE. 

For comparison, the conventional IPRs defined via solely right eigenvectors $\psi_j$ (i.e., with $\rho_j=|\psi_j|^2$) are also shown in \figref{Fig2}(b), where black and gray curves represent the edge and bulk states, respectively. Unlike the bi-IPRs, the conventional IPRs do not show a clear-cut distinction between the edge and bulk states, since the bulk right eigenvectors are also localized due to NHSE. The transitions of the black curve occur when $u_-=s$, namely at $k_x \approx 0.46\pi$ and $1.54\pi$ (corresponding to {\it{``R''}} in \figref{Fig2}(a)), which differ from the bulk-edge degeneracies at $k_x \approx 0.49\pi, 1.51\pi$. Thus the bi-IPR is a good measure to characterize the bulk and edge states. 

The red (blue) curve reaches unity at those $k_x$ values that correspond to EPs (NDPs) in \figref{Fig2}(a), namely where $u_\pm=0$ ($s=0$), because of complete localization of $\rho_j$ at one site. The black curve reaches unity when $u_-=0$ and $s=0$ due to complete localization of $\psi_j$. 

The rationale behind the non-Bloch approach is to find a substitute NHSE-free system that shares the same OBC spectrum with the original system so that the BBC can be established. Following the non-Bloch approach \cite{yao2018Edge,yao2018Chern} with $\beta=|\beta| e^{i\theta}$ substituted for $e^{ik_y}$ in \eqnref{eq:Hk}, the GBZ is found to be a circular loop with radius 
\eq{ |\beta|=\sqrt{|u_-/u_+|} 
\label{GBZ}
}
for given $k_x$. The 2D GBZ is then a torus parameterized by $k_x$ and $\theta$, both with the range $[0,2\pi]$. By the replacement $e^{ik_y} \to |\beta| e^{i\theta}$ in \eqnref{eq:Hk}, we obtain the non-Bloch Hamiltonian 
\eq{\label{eq:Hkp}
H(\vect{k}^{\prime}) =
\left( \begin{array}{ccccc}
m & u_+ + s |\beta|^{-1}e^{-i \theta} \\
u_- + s |\beta| e^{i \theta} & -m
\end{array} \right),
} 
where $\vect{k}^{\prime}=(k_x, \theta)$, which has the same bulk eigenspectrum as the OBC system associated with \eqnref{eq:Hk}. 

All bulk eigenvectors are delocalized at $k_x=2\pi/3$ and $4\pi/3$, namely when $|\beta|=1$, because $u_+=-u_-=g$ and the NHSE is absent. The bulk right eigenvectors are localized at the lower boundary when $|\beta|<1$, that is, when $k_x\in (2\pi/3,4\pi/3)$, and they are localized at the upper boundary for $k_x\in[0,2\pi]-[2\pi/3,4\pi/3]$. The opposite goes for the left eigenvectors and therefore the combination of the left and right eigenvectors suppresses the NHSE. 

Equation \eqref{eq:Hkp} is not well defined at $k_x=k_1^+$, $k_2^+, k_1^-, k_2^-$, which are associated with EPs in \figref{Fig2}(a), because either off-diagonal element diverges due to $|\beta| = \infty$ or $|\beta|^{-1} = \infty$. However, if choose a different unit cell in \figref{Fig1}(a) so that $t\pm g$ become intercell hoppings, we obtain the non-Bloch Hamiltonian with a different form: 
\eq{\label{eq:hkp}
h(\vect{k}^{\prime}) =
\left( \begin{array}{ccccc}
-m & s+u_- |\beta|^{-1}e^{-i \theta} \\
s+ u_+ |\beta| e^{i \theta} & m
\end{array} \right).} 
Equation \eqref{eq:hkp} is free of divergence problem because $u_- |\beta|^{-1}=\sgn{(u_-)} \sqrt{|u_- u_+|}$ and $u_+ |\beta|=\sgn{(u_+)} \sqrt{|u_- u_+|}$ and both vanish at $k_x=k_1^+,k_2^+,k_1^-,k_2^-$. And there is no NHSE for $h(\vect{k}^{\prime})$ since $u_- |\beta|^{-1} = \pm u_+ |\beta|$ \cite{gong2018PRX}. Equation \eqref{eq:hkp} without $|\beta|$ and $|\beta|^{-1}$ is the corresponding Bloch Hamiltonian $h(\vect{k})$. 

Equations \eqref{eq:Hkp} and \eqref{eq:hkp} share the same non-Bloch bands, which are shown by red curves in \figref{Fig2}(c), with solid (dashed) curves for the real (imaginary) parts. For comparison, the Bloch bands $E_\pm (\vect{k})$ from $H(\vect{k})=H(k_x,k_y )$ are also shown with black (gray) curves for the real (imaginary) parts. The horizontal axis represents $\vect{k}^{\prime}=(k_x,\theta)$ and $\vect{k}=(k_x,k_y )$ for the non-Bloch and Bloch bands, respectively. Two band structures almost coincide in some regions, but they show notable differences at $k_x= k_1^+$ and $k_1^-$. This is expected because the NHSE is most prominent near the EPs. In particular, the $Re[E_\pm (\vect{k}^{\prime})]$ bands exhibit sharp cusps when $k_x=k_1^+$ and $k_1^-$. Taking $\theta=0$ (i.e., along $\Gamma X$) for example, the cusps are caused by the presence of the term $f=\sqrt{|u_- u_+|} [\sgn(u_-)+\sgn(u_+)]$ in $E_\pm (\vect{k}^{\prime} )=\pm \sqrt{m^2+s^2+u_- u_+ +sf}$. 

Unlike $H(\vect{k})$, the non-Bloch system $h(\vect{k}^{\prime})$, as well as $H(\vect{k}^{\prime})$, is free of the NHSE, which is essential to the non-Bloch approach. Since the Chern number is independent of the choice of unit cell,
$h(\vect{k}^{\prime})$ can serve as the bulk system in establishing the BBC. The validity of $h(\vect{k}^{\prime})$ is further confirmed by the fact that the OBC bulk spectrum in \figref{Fig2}(a) coincides with the projection bands computed from $h(\vect{k}^{\prime})$. 

We define the non-Bloch Chern number as an integral over the GBZ \cite{yao2018Chern,shen2018Topological}, 
\eq{C_n=\frac{1}{2\pi}\int_{GBZ}\epsilon_{ij}B^n_{ij}d^2\vect{k}^\prime, 
\label{Chern}
}
where $n=\pm$, $i,j=x,y$, and $\epsilon_{xy}=-\epsilon_{yx}=1$, and $B_{ij}^n=i \langle \partial_i \varphi_n (\vect{k}^\prime) | \partial_j \psi_n (\vect{k}^\prime) \rangle $ with $\partial_x\equiv\partial_{k_x}$ and $\partial_y\equiv\partial_\theta$ \cite{yao2018Chern,shen2018Topological}. The normalization condition $\langle\varphi(\vect{k}^\prime ) | \psi(\vect{k}^\prime) \rangle=1$ is assumed for the left and right eigenvectors of $h(\vect{k}^{\prime})$. $C_\pm=\pm 1$ is obtained for the $E_\pm (\vect{k}^\prime)$ bands \cite{SeeSupp}. The BBC is now established that the number of edge bands, identified by the biorthogonal density and IPR, is dictated by the non-Bloch Chern numbers, in a similar manner as the conventional BBC. 
\begin{figure}[tbp]
\includegraphics[clip, width=\columnwidth, angle=0]{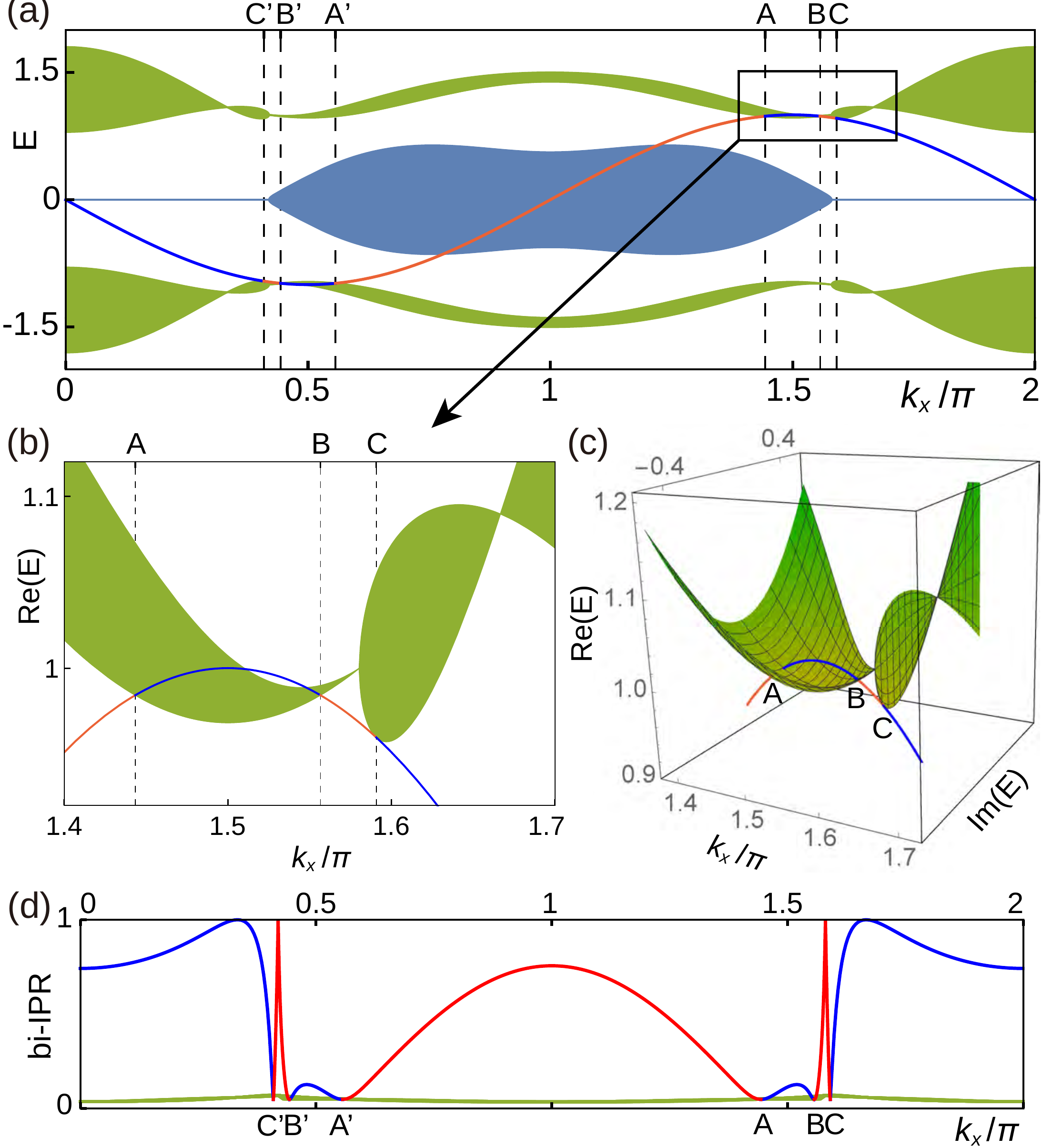}
\caption{(a) The band structure of the OBC system with $\delta=\Delta=1$, $t=0.5$ and $g=0.75$. The green (navy) curves denote $Re(E)$ and $Im(E)$ of bulk bands, and the blue/red curve is the edge band. (b) The zoomed view of (a). (c) The crossing degeneracies {\it{A}}, {\it{B}} and tangential degeneracy {\it{C}} occur at the intersections and tangential point, respectively, of the edge-band curve with the surface that represents the complex bulk eigenvalues. (d) The bi-IPR of the bulk states (green) and edge states (blue/red). 
\label{Fig3}}
\end{figure}
\subsection*{C. Topological phase with crossing degeneracies} 
Next we investigate another nontrivial case with $g$ increased to $0.75$. The spectrum of the system with OBC is shown in \figref{Fig3}(a). The real (imaginary) part of bulk bands is denoted by the green (navy) curves, and the edge band $E=-\Delta\sin k_x$ is colored in red/blue for states with bi-IPRs localized at the lower/upper boundary. The zoomed view of the band structure is shown in \figref{Fig3}(b), where the dashed lines {\it{A}} and {\it{B}} mark an unusual type of bulk-edge degeneracies, which we coin as crossing degeneracies, and the line {\it{C}} marks the tangential degeneracy. 
The crossing degeneracies {\it{A}}, {\it{B}} and tangential degeneracy {\it{C}} manifest themselves clearly as the intersections and tangential points, respectively, of the edge-band curve with the surface that represents the complex bulk eigenvalues, as shown in \figref{Fig3}(c). 
Their mirror images {\it{C'}} ({\it{A'}}, {\it{B'}}) in \figref{Fig3}(a) also correspond to tangential (crossing) degeneracies. 

All bulk-edge degeneracies become exact when $N_y \to \infty$ and can be located by $u_+ u_-=\pm s^2$, where ``$\pm$'' correspond to the two cases when $h(\vect{k}^\prime)$ is Hermitian and non-Hermitian, respectively. Therefore bulk eigenvalues are purely real around {\it{C}} (and {\it{C'}}), which manifest as a vertical flat suface in \figref{Fig3}(c), and the degeneracies there are forced to be {\it tangential}. In contrast, the bulk spectrum around {\it{A}}, {\it{B}} (and also {\it{A'}}, {\it{B'}}) contains both complex and purely real eigenvalues and manifests as a curved surface shown in \figref{Fig3}(c), which allows {\it{crossing}} degeneracies with the edge band. The crossing degeneracies are thus intrinsically non-Hermitian and do not exist in Hermitian systems.

The bi-IPRs are shown in \figref{Fig3}(d), where the red/blue curves denotes edge states that are localized at the lower/upper boundary, and the green curves represent the bulk states. All the delocalization points of the edge-band bi-IPRs correspond to the bulk-edge degeneracies in \figref{Fig3}(a). The edge-band bi-IPRs reach the largest value of unity at $k_x \approx 0.42\pi, 1.58\pi$ and also at $k_x=5\pi/3$, where EPs and NDPs occur in \figref{Fig3}(b), due to the single-site confinement of the biorthogonal density \cite{SeeSupp}. 

The edge band $E=-\Delta\sin k_x$ is divided into six segments by the bulk-edge degeneracies (i.e., delocalization points) {\it A},{\it B},{\it C},{\it A}',{\it B}',{\it C'}. Only the two segments {\it A}'{\it A} and {\it C}{\it C}' traverse the gap and connect the valence band to the conduction band, and are thus topologically robust. In contrast, the segments {\it AB} and {\it B}'{\it A}', each of which are pinned by crossing degeneracies in the same bulk band, can be eliminated by tuning parameters without closing the gap so that the bands resembles the $g=0.25$ case (\figref{Fig2}) with only tangential degeneracies. The case of \figref{Fig3} therefore lies in the same nontrivial phase as \figref{Fig2}, as is validated by $C_\pm=\pm 1$ calculated for the non-Bloch bands. 
\subsection*{D. Phase diagram}
The Hermitian systems ($h(\vect{k}^\prime)$ with $g=0$) is gapless when $\delta=0$. The presence of hoppings $\pm\delta/2$ with $\delta \ne 0$ opens a gap with real-valued bands $E_\pm (\vect{k}^\prime )$ and $C_\pm=\pm 1$. At $g=0$, the $Im(E_\pm)$ bands are identically zero. Assuming a large $g$, \eqnref{eq:hkp} becomes $h(\vect{k}^{\prime})=-m\sigma_z+(s+ig^{\prime}\sin\theta)\sigma_x-ig^{\prime}\cos{\theta}\sigma_y$ with $g^{\prime}=\sqrt{g^2-u^2}>0$ and $\sigma_i$ denoting Pauli matrices, and the eigenvalues become 
$E_{\pm}(\vect{k}^{\prime})\approx\pm(ig^{\prime}+s\sin\theta)$
, the real parts of which are always gapless at $\theta=0$ and $\pi$ \cite{SeeSupp}. It is expected that $C_\pm=0$ since the time-reversal breaking parameter $\Delta$ becomes negligible when $g\to\infty$. As $g$ is increased from $0$ to $\infty$, the non-Bloch bands therefore must undergo a topological transition from $C_\pm=\pm 1$ to $C_\pm=0$, and two non-Hermitian phase transitions that correspond to the gap-closing of $Re(E)$ bands and the gap-opening of $Im(E)$ bands, respectively. The non-Hermiticity $g$ is thus expected to induce rich features to the phase diagram. 

\begin{figure}[htbp]
\includegraphics[clip, width=\columnwidth, angle=0]{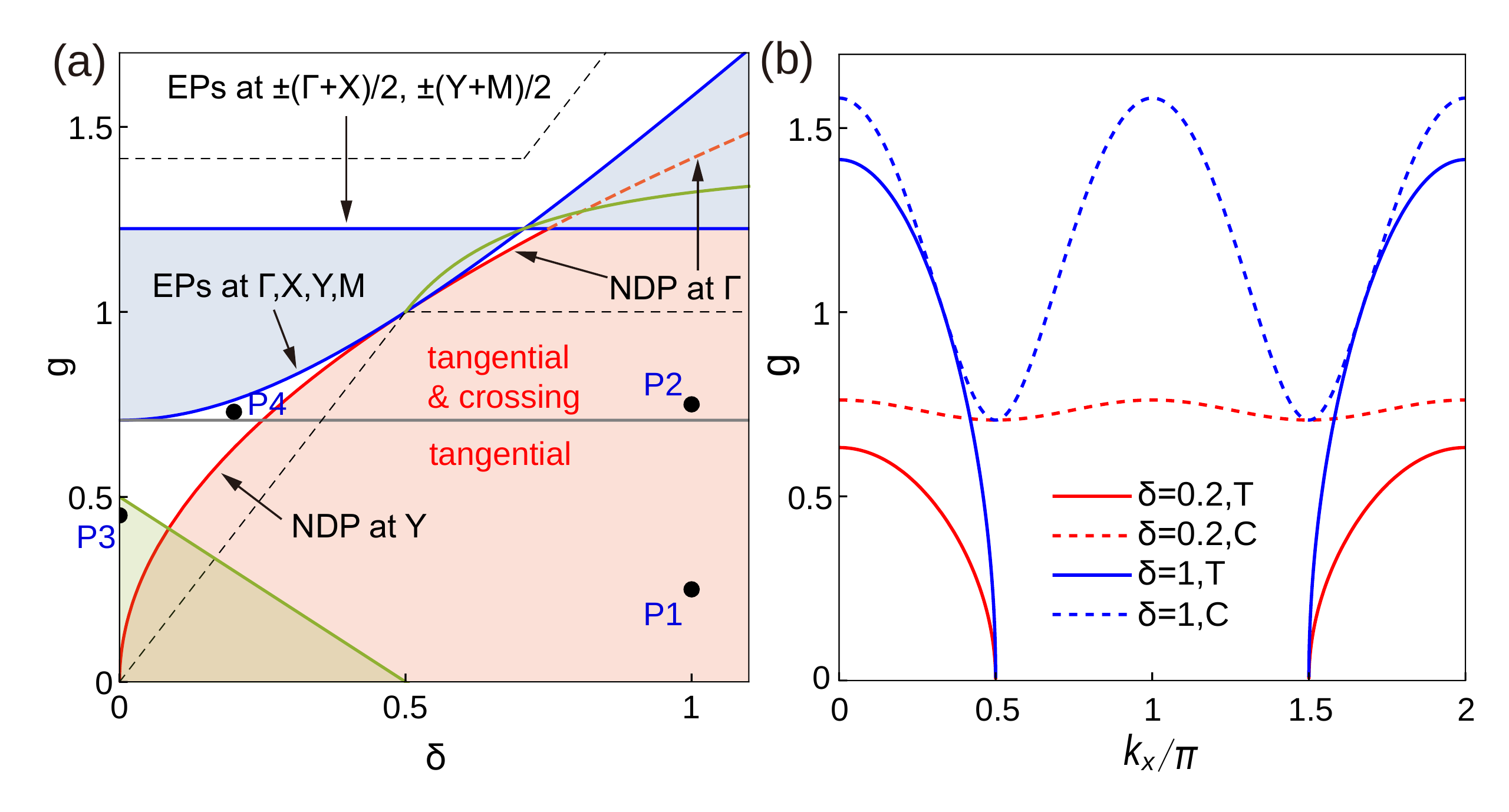}
\caption{(a) The phase diagram in the $\delta g$-plane. The nontrivial/trivial/gapless phase region is colored in red/white/blue. The green region supports purely real spectrum. The crossing bulk-edge degeneracies occur in the region between the gray line and the blue hyperbola. The green curve corresponds to the occurrence of EP lines. (b) The evolution of the tangential (T) and crossing (C) bulk-edge degeneracies with increasing $g$ for $\delta=0.2$ and $\delta=1$. 
\label{Fig4}}
\end{figure}

Without loss of generality, we assume $\Delta=1$ and $t=0.5$. The phase diagram in the $\delta g$-plane is shown in \figref{Fig4}(a), where the red, white and blue regions represent nontrivial gapped ($C_\pm=\pm 1$), trivial gapped ($C_\pm=0$) and gapless phases, respectively. The red and two blue phase boundaries respectively correspond to the aforementioned topological and two non-Hermitian phase transitions, and can be derived analytically by requiring $|E(\vect{k}^\prime)|=0$ \cite{SeeSupp}. 

The system is gapless at $(\delta,g)=(0,0)$, and the hoppings $\pm\delta/2$ with $\delta \ne 0$ are necessary to open a non-trivial gap when $g=0$. In contrast, a nonzero $g$ opens a trivial gap when $\delta=0$. There must then be the red phase boundary curve in \figref{Fig4}(a) that corresponds to topological transitions. It is a parabola $g=\sqrt{4t\delta}=\sqrt{2\delta}$ and corresponds to gap closing and reopening accompanied by an NDP (Dirac point) at $\Gamma$ or $Y$ \cite{SeeSupp}, reminiscent of Hermitian topological transitions.

The blue phase boundaries consist of a hyperbola $g=\sqrt{2\delta^2+2t^2}=\sqrt{2 \delta^2+1/2}$ and a line $g=\sqrt{\Delta^2+2t^2 }=\sqrt{6}/2$ without $\delta$ dependence, and they are associated with EPs at $\Gamma,X,M,Y$ and $\pm \frac{\Gamma+X}{2}, \pm\frac{Y+M}{2}$, respectively \cite{SeeSupp}. They correspond to non-Hermitian phase transitions at EPs, in contrast to the topological transitions associated with NDPs. A topological invariant $\nu$ called discriminant number, 
\eq{\nu\left(\kp_{EP}\right) = \frac{1}{2\pi i}\oint_{C\left( \kp_{EP}\right)}{d\kp\cdot \nabla_{\kp}}\ln{\mathcal{D}_{f}\left( \kp \right)},
}
can characterize the topology of the discrete EPs \cite{xiao2020Lieb,yang2021Doubling,SeeSupp}, where $C\left( \vect{k}_{EP}^{\prime} \right) $ denotes a loop that encircles $\vect{k}_{EP}^{\prime}$ and $\mathcal{D}_{f}\left( \kp \right)$ denotes the discriminant of the characteristic polynomial $f_E(\kp)=\det[E-h(\kp)]$. 

At the blue phase boundaries, there are $4$ unstable EPs characterized by $\nu=0$ and they each get gapped or split into two EPs with $\nu=\pm 1$ when leaving the boundaries \cite{SeeSupp}. When $(\delta, g)$ is tuned from the blue hyperbola to the blue line, the $8$ EPs with $\nu=\pm 1$ move from $\Gamma, X$ to $\pm \frac{\Gamma+X}{2}$, or from $M,Y$ to $\pm\frac{Y+M}{2}$, with their locations given by $\vect{k}^\prime=(k_x, \theta)$ where $k_x$ satisfies $\cos{(2k_x)}=\frac{g^2-\delta^2-\Delta^2}{\delta^2-2t^2}$ and $\theta=0,\pi$ \cite{SeeSupp}. During the evolution, the discrete EPs remain stable due to the topological protection by nonzero discriminant numbers $\nu=\pm 1$ \cite{SeeSupp}. Therefore the gapless phases have a topological origin. 

Since $\sqrt{4t\delta} \le \sqrt{2t^2+2\delta^2}$ with the equality satisfied only when $\delta=t$, the red parabola (topological transition) lies below the blue hyperbola and they must touch at $\delta=t$.

As $g$ is increased, the lower (upper) blue boundary corresponds to the gap closing (reopening) in the $Re(E_\pm)$ ($Im(E_\pm)$) bands, confirming our analysis. Their crossing at $(\delta,g)=(\sqrt{2}/2,\sqrt{6}/2)$ corresponds to the simultaneous gap closing and reopening. For $\delta<3/4$ (i.e., $\sqrt{2\delta}<\sqrt{6}/2$), the mere increase of non-Hermiticity $g$ thus not only induces two non-Hermitian phase transitions, but also a topological transition, which is quite unusual. For $\delta>3/4$, a gapless range lies between the trivial and nontrivial phases. 

The touching point $(\delta, g)=(0.5,1)$ signifies the switching of the NDP from $Y$ to $\Gamma$ at the topological (red) phase boundary \cite{SeeSupp}. In addition to EPs at $X,M$, a NDP nodal line occurs at $\Gamma Y$ for $(\delta, g)=(0.5,1)$, because $m=s=u_-=0$ and $h(\vect{k}^\prime)$ becomes a zero matrix. Along the green trajectory $g=\sqrt{2-(t/\delta)^2 }$ in the gapless phase, two EP lines arise from the splitting of the NDP nodal line and occur at $k_x=\pm\cos^{-1}(t/\delta)$ in the $k_x\theta$-plane. The EP lines are associated with $s=0$. When $(\delta, g)$ is tuned to cross the green trajectory, $4$ of the $8$ discrete EPs change to the EP lines and then back to $4$ discrete EPs, accompanied by a switch in their charges, i.e., $\nu=\pm 1 \to \nu=\mp 1$ \cite{SeeSupp}. 

The green region $g+\delta \le t$ supports purely real OBC spectrum since $h(\vect{k}^\prime)$ becomes Hermitian due to $u_+ \ge u_- \ge 0$ for any $\vect{k}^\prime$. 

The crossing degeneracies occur in the region between the gray line $g=\sqrt{2}/2$ and the blue hyperbola, and tangential degeneracies occur in the lower region bounded by the red parabola. The points P1 and P2 separated by the gray line correspond to the two typical cases shown in Figs. 1 and 2, respectively. Similarly, P3/P4 represents two typical trivial phases with/without crossing degeneracies \cite{SeeSupp}. 

For comparison, the phase boundaries determined from $H(\vect{k})$ are shown as black dashed lines in \figref{Fig4}(a), which both correspond to non-Hermitian phase transitions and deviate drastically from the (correct) red/blue phase boundaries derived from $h(\vect{k}^\prime)$. The red/blue boundaries are circumscribed by the black dashed lines, which arises from the fact that the OBC bulk spectrum is enclosed by the PBC spectrum in the complex plane \cite{okuma2020Origin,wang2019Transition}, that is, $\sigma(OBC)$ may still be gapped when $\sigma(PBC)$ becomes gapless. 
\subsection*{E. Evolution of bulk-edge degeneracies}
Next we show the evolution of bulk-edge degeneracies and their relevance to phase transitions. In \figref{Fig4}(b), we plot the trajectories of $k_x$ values that correspond to the tangential (solid) and crossing (dashed) degeneracies as $g$ is increased. The solid and dashed curves are given by $g=\sqrt{4t\delta\cos{k_x}}$ and $g=\sqrt{2t^2+2\delta^2\cos^2{k_x}}$, which are derived from \eqnref{delocalize}, and they are shown as red (blue) curves for $\delta=0.2$ ($\delta=1$). $\Delta$ and $t$ remain the same as in \figref{Fig4}(a). The cutoff $g$ value for the tangential degeneracies at $k_x=0$, namely $g=\sqrt{4t\delta}$, gives the red phase boundary in \figref{Fig4}(a), because the bulk bands that come in $\pm E_B(k_x)$ pairs become gapless noting that their degeneracy with the edge band $E=-\Delta\sin k_x$ at $k_x=0$ implies $\pm E_B=0$, namely gap closing. 

The crossing degeneracies emerge when $g$ reaches a threshold of $\sqrt{2}/2$ for any value of $\delta$, indicating their non-Hermitian origin. Such degeneracies disappear at $g=\sqrt{2\delta^2+1/2}$, which coincide with the blue hyperbolic phase boundary in \figref{Fig4}(a). The domain of $g$ pinned at $k_x=0$ by the two red curves in \figref{Fig4}(b) is thus the range of the trivial phase for $\delta=0.2$ in \figref{Fig4}(a). $g=2t=1$ derived from $u_+u_-=\pm s^2=0$ corresponds to the coincidence between the tangential and crossing degeneracies at two $k_x$ values where $\cos k_x=t/\delta$ for $\delta \ge t$, as is shown in \figref{Fig4}(b). Their coincidence corresponds to $h(\vect{k}^\prime)=-m\sigma_z$ and therefore $n_y$-fold degeneracies at $E=\pm m$ in the ribbon bands. 
\section*{III. Conclusion}
We establish the BBC for a prototypical non-Hermitian Chern insulator model with NSHE by combining the non-Bloch approach and biorthogonal approach, both aiming at suppressing the NHSE. Specifically, a Chern number is defined for the non-Bloch bands and we use the biorthogonal IPRs instead of the wave functions to distinguish between the edge and bulk bands, utilizing the fact that the bulk-edge degeneracies coincide with the delocalization of edge states in terms of the biorthogonal density. The system undergoes one topological transition and two non-Hermitian phase transitions when only a single parameter is tuned. That leads to a rich phase diagram which contains nontrivial gapped, trivial gapped and gapless phases. The gapless phase is topologically protected due to the stability of the EPs ensured by nonzero discriminant numbers. The presence of crossing degeneracies which are intrinsically non-Hermitian enriches the properties in all three phases.
\section*{Acknowledgements}
We thank Professor Zhao-Qing Zhang for many helpful suggestions. This work is supported by Research Grants Council (RGC) Hong Kong through grant 16303119 and 16307420. We thank the reviewer for bringing up interesting questions on the delocalization.  
%
%
%
\nocite{xiao2019AnisotropicEP}
\end{document}